\documentclass{article}
\usepackage[utf8]{inputenc}
\usepackage{authblk}
\usepackage{amsmath}
\usepackage{amssymb}
\usepackage{graphicx}
\usepackage{float}
\usepackage{color} 
\usepackage[T1]{fontenc}

\providecommand{\keywords}[1]
{
  \small	
  \textbf{\textit{Keywords---}} #1
}

\title{Waves of desertification in a competitive ecosystem}

\author[1]{Y. C. Daza C. \thanks{yudy.daza@cab.cnea.gov.ar}}
\author[1]{M. F. Laguna \thanks{lagunaf@cab.cnea.gov.ar}}
\author[3,4]{J. A. Monjeau \thanks{amonjeau@fundacionbariloche.org.ar}}
\author[1,2]{G. Abramson \thanks{abramson@cab.cnea.gov.ar}}

\affil[1]{\small{Centro At\'{o}mico Bariloche and CONICET, R8402AGP Bariloche, 
Argentina}}
\affil[2]{\small{Instituto Balseiro, R8402AGP Bariloche, Argentina}}
\affil[3]{\small{Fundaci\'{o}n Bariloche and CONICET, R8402AGP Bariloche, Argentina}}
\affil[4]{\small{Laborat\'{o}rio de Ecologia e Conserva\c{c}\~{a}o de 
Popula\c{c}\~{o}es, Departamento de Ecologia, Universidade Federal do Rio de 
Janeiro, Rio de Janeiro, Brasil}}

\begin{document}

\maketitle

\begin{abstract}
We study a mathematical model inspired by a common scenario in northern 
Patagonia consisting of humans and their livestock living together with 
native wildlife. The main production system in this area is sheep farming, 
which due to wrong historical management has led to desertification of the 
habitat, with impact on both native wildlife and livestock. Particularly the 
largest native herbivore, the guanaco, has reflected in their 
numbers this alteration induced in the environment. 
We analyze a mathematical model that captures the main 
characteristics of the interaction between sheep and guanaco: the 
hierarchical competition and the advantage granted by humans to the herds, and 
also incorporates a dynamic for the habitat. Using the 
metapopulation formalism, the trophic web of two herbivores is extended over a 
patchy landscape that considers two characteristic times for the dynamics of 
the resource. 
Our study stresses the dependence of the metapopulations dynamics on the 
recovery time of the resource. 
These results are backed up by a deterministic mean field 
model which shares some similarities with the stochastic and spatially 
extended one.
We find different regimes depending on the parameters considered: coexistence 
of both species, survival of a single species and extinction of the other, and 
extinction of both. Remarkably, in some regions of parameters space we detect 
the presence of periodical spatio-temporal patterns, with persistent 
oscillations of constant amplitude.
Based on these results, we perform a characterization of the observed scenarios 
in order to gain insight about the system.
\end{abstract}


\date{\today}

\section{Introduction}

This model is based on a true story.

Indeed, the mathematical model we have developed is inspired by the history of 
the environmental devastation of Patagonia. Located in the southernmost 
tip of South America, in Argentina, the Patagonian steppe is an emblematic 
landscape \cite{SMAyDS}. The extra-Andean Patagonia comprises  c. 750~000 
km$^2$ of arid and semiarid lands where extensive sheep ranching is the 
predominant land use \cite{VonThungen,Nabteetal}.  The unsustainable 
management of sheep farming is as old as the expulsion of native peoples by the 
Argentine army, just over 100 years ago, replacing a hunter-gatherer system of 
thousands of years by a system of private lands, subdividing a landscape, 
previously continuous, in ranches in turn divided internally by wire fences 
\cite{Delrio}.

The basis of the subsistence economy of the original peoples of Patagonia was 
the use of native herbivores, especially the guanaco (\emph{Lama guanicoe}) 
\cite{Munoz}. 
The guanaco has coevolved in the arid ecosystems, minimizing the environmental 
damage in the construction of its niche \cite{Puigetal}. As groups of guanacos 
constantly 
move in search of good pastures, there is a dynamic in which palatable and less 
palatable species alternate their abundance over time and space, allowing a 
sustainable foraging of the resource (when it occurs in natural conditions), 
and the concomitant  control by predators, such as the puma (\emph{Puma 
concolor}). The Tehuelche 
people built their niche learning to follow immense groups of guanacos, of up 
to 300~000 individuals, in their seasonal migrations, a nomadic system that 
subsisted in a sustainable way for at least 6000 years \cite{Musters}. 

After the so-called ``conquest of the desert'' at the end of the nineteen 
century \cite{Delrio}, sheep and 
guanaco began to coexist as competitors for the forage resource 
\cite{Baldietal}. 
In the early days of sheep farming, with large areas and low population 
density, a system similar to that of nomadism was common. Shepherds on 
horseback 
carried the flocks through vast extensions in search of good pastures 
\cite{Fernandezetal}. The main movement, as in the case of guanacos and 
Tehuelches, occurred 
between winter and summer. In summer high places were used letting the low 
valleys rest, which were used in winter. This system still prevails in certain 
areas of Patagonia, such as Payunia in the provinces of Neuquén and Mendoza.

In another modality, the holistic management, the ranches are exploited in a 
way that maximizes the harness of the biomass by the flocks \cite{Savoryetal}. 
The 
most common modality is, however, livestock management in paddocks 
\cite{Hobbsetal}, where the aim is to maximize short-term gain 
\cite{Ottichiloetal}. Obviously, 
as the wildlife is out of the market, it is perceived as an impediment to the 
maximization of profits, which is the source of conflicts between productive 
activities and the conservation of native species \cite{Baldietal, Prins}. 
Land use in Patagonia often implies negative consequences for wildlife  and 
these are 
particularly evident when the removal of certain species is supposed to 
increase landholders' income or reduce production costs \cite{Chaikinaetal}. 
This conflict 
is exacerbated in arid rangelands or in periods of drought, where forage and 
water availability show high variability in space and time \cite{Wrobeletal}, 
intensifying 
wildlife prosecution \cite{Nabteetal}. Previous studies have addressed that 
due to direct and indirect competition, wild herbivores abundance is inversely 
correlated 
with livestock density \cite{Baldietal, Pedranaetal}. In particular, the 
interspecific competition of the guanacos with livestock lies in the fact that 
both 
herbivores are generalists and their diets overlap up to 83\% \cite{Puigetal}, 
and landowners perceive a decrease in carrying capacity when 
guanaco 
abundance increases \cite{Nabte2010}. 

Given that the inadequate management of livestock is the general rule (having 
more animal load than the carrying capacity) the deterioration of the habitat, 
the loss of the species of greater nutritional value and the extinction of 
wildlife are the most common consequences of the current situation 
\cite{Golluscioetal, Fuhlendorfetal}. 
%Additionally, the gradual subdivision of properties due to the inheritance of 
%ranches among many heirs and other economic factors, implies an 
%increase in disturbances due to management, habitat fragmentation and the 
%decrease of environmental heterogeneity, with negative consequences for 
%biodiversity 
%and ecosystem functioning \Citep{duToit}.

Depending on the use of the land, the balance between wildlife and livestock 
use can be very different. In the few protected areas of Patagonia, which 
occupy only 4.62\% of its surface \cite{Nabteetal} a focus on 
conservation is expected. In contrast, landholders 
relying on entirely productive activities
often have no reason to offset the costs derived from wildlife tolerant 
practices \cite{Victurineetal}. Therefore, when resources are limited, or 
international 
prices put pressure on the landholders, the competition of guanaco with sheep 
becomes very significant. In such circumstances it is expected that the main 
motivation of the ranchers 
is to solve the problem by exterminating or driving away wildlife, since it 
does not bring them any benefit \cite{Nabte2010}. It is necessary to 
understand 
how 
these anthropic factors affect decision making regarding the management of 
sheep farming and its coexistence with wildlife. In most ranches, top predators 
(\emph{P.
concolor} and \emph{Lycalopex culpaeus}) and herbivores (\emph{L. guanicoe}, 
\emph{Pterocnemia pennata}) have been locally extirpated by hunting or scare to 
reduce economic losses \cite{Nabteetal, Nabte2010} 
 and to feed domestic dogs and cats \cite{Novaro, Gallardoetal, DeLucca, 
Travainietal}. 

Despite the fact that the guanaco is highly fit to inhabit the Patagonian 
desert, its population density has diminished dramatically in historical times 
by having to share its natural environment with humans. If there were no human 
intervention, the guanaco herds, due to their size and natural history, would  
competitively impose on herds of sheep in the territorial struggle for 
resources such as food or water. But the current scenario tells an unnatural 
story. The flocks along with their guards (human, guns, dogs, fences: the 
livestock system) make 
difficult the survival of guanaco populations when the better pastures and 
wetlands are monopolized by human activities, and the populations are killed or 
displaced to marginal places with poor plant cover and nutrient deficiency; the 
same goes for other wildlife species, such as native deer 
\cite{flueck2006,flueck2011}. Then, 
the places occupied by the livestock suffers degradation due to the intrinsic 
foraging characteristics of the sheep and the wrong management plans that 
privileges short-term gains rather than long-term sustainability. As a result, 
many ranches were desertified and abandoned, ruining the natural capital for 
humans \cite{andrade2013} and wildlife \cite{bonino2005}.

This introductory context is the basis for the assumptions we make in the 
dynamical model presented here. In a previous article we explored other aspects 
of the same problem with a similar model \cite{laguna2015}. 
In this work we analyze an ecological system inspired in the hierarchical 
competitive interaction of guanaco and sheep. The problem is addressed in the 
framework of metapopulations, which provides the  possibility of a mechanistic 
understanding of the biological consequences of habitat loss and fragmentation 
\cite{hanski1998, hanski2017}.
The occupation of patches is driven by ecological processes that occur in a 
stochastic way (extinction  and colonization) and also are modulated by the 
patch 
state, which in turn is determined by the dynamics of the competitors. We show 
that an interesting dynamics arises, showing scenarios of coexistence, 
extinction and metapopulation waves.

\section{Mathematical model}

We consider two species in hierarchical competition that affects 
an extended resource. 
As we said, the guanaco is competitively superior to the sheep due to its 
character and size, but human intervention makes the sheep better moving around 
and colonizing new patches. These features are included in the model through 
the rules of interaction and the value of the parameters chosen for each 
species.

Besides, the result of grazing is different for both species. The guanaco has 
co-evolved with the flora they consume and, under natural conditions, 
there is a spatio-temporal dynamics of abundance among species of different 
palatability, allowing a sustainable consumption over time. For this reason, in 
the model, the impact of the guanaco on the habitat will be considered as 
negligible. 
The case of sheep is very different, since overgrazing can cause a strong 
impact on the resource, affecting both herbivores in a negative feedback. In 
such a context, the key factor that determines the survival of any species is 
the dynamics of the spatio-temporal structures that emerge from the ecological 
interactions. 

%Metapopulation models are methods for the study of the ecological relations 
%inscribed in a spatially structured landscape. In this approach the landscape 
%or the resource is fragmented in patches, which is a reasonable idealization 
%for some situations in which the species are confined and moves from one place 
%to another by migrations. 
%This characteristic make this approach a good choice for understand the 
%spatial dynamics of the resource under the grazing of the sheep and guanaco 
%populations and to understand how the occurrence of the species is affected by 
%the dynamics of the patches. 

\subsubsection*{The metapopulation model for the two herbivores}

The classical approach of metapopulations consists of a system formed of local 
populations of a given species, located in specific places called patches 
\cite{levins1971,hanski1994}. It is assumed that, for the survival of a local 
population, the patch must provide sufficient conditions to maintain and 
sustain them. Then, the patches are modeled like places suitable or unsuitable 
for be the habitat of a given species. 
The metapopulations are driven by ecological processes like death, 
colonization and interspecific competition. 
Competition can be captured by the model making the superior species (guanaco) 
and the inferior one (sheep) unable to share the same 
patch, as the inferior 
herbivore will be repealed from any patch already occupied by the superior one, 
or could be expelled if a superior migrates to the patch that is occupying.
Colonization, on the other hand, will be determined by the human intervention. 
Although the guanaco is capable for surviving in regions where the sheep can 
not, the presence of the human makes the balance inclined towards the sheeps, 
as herds are protected and moved from one place to another. In this work we 
consider the colonizing capability of sheeps as a parameter always greater than 
the colonizing capability of the guanacos.
Finally, deaths have to be understood as extirpation 
from a patch. This process can occur by a diversity of 
mechanisms, like death of the local population that occupy the patch, or 
migration to another patch, but the focus is always on the property of 
\emph{occupation} of a patch. The extinction of a species in this context 
corresponds to their disappearance from all the patches.

We model the ecological processes that drive the dynamics just described as 
stochastic processes, in order to reflect the demographic fluctuations 
\cite{hanski1998}. The dynamical variables are $x_1$ and $x_2$, the fraction 
of patches occupied by the superior and the inferior competitor, respectively. 
Local extinctions occurs at random with probabilities $e_1$ and $e_2$. 
Colonization takes place in the model according to the availability of patches 
in the immediate neighborhood of an occupied one. It is also a stochastic 
process characterized by $c_1$ and $c_2$. The availability of patches reflects 
both the dynamics of the resource (to be specified below) and the occupation by 
other populations. Given the hierarchical nature of the competitive interaction 
we consider two cases: 1) a patch can be occupied by the superior herbivore if 
it is free of their own, regardless of the presence of the inferior competitor; 
2) a patch can be occupied by the inferior herbivore if it is free of both 
species. Further, in a patch occupied by both species, the inferior competitor 
can be displaced by the superior one with probability $c_1$. Together, these 
rules characterize the asymmetric (hierarchical) competition. A detailed 
discussion can be found in our previous works \cite{laguna2015,abramson2017}.

\subsubsection*{Dynamics of the habitat}

As stated above, a patch can be in one of two states: empty or occupied. 
This binary state would fluctuate at each patch according to 
the dynamics of the metapopulations. Besides, the resources at each patch can 
change according to a number of processes, such as the renewal of the pastures 
according to environmental factors like temperature, precipitation, etc., as 
well as 
the consumption by both species. Depending of the pressure of grazing the 
resource, $h$, may remain sustainable or become depleted. If the continuous 
occupation by the inferior competitor depletes $h$ at a patch, that patch 
becomes unsuitable for colonization for both $x_1$ and $x_2$. In an optimistic 
scenario, we can assume that the resource has the power to recover the 
pre-occupation level. 

\begin{figure}[t]
 \centering
 \includegraphics[scale=0.25]{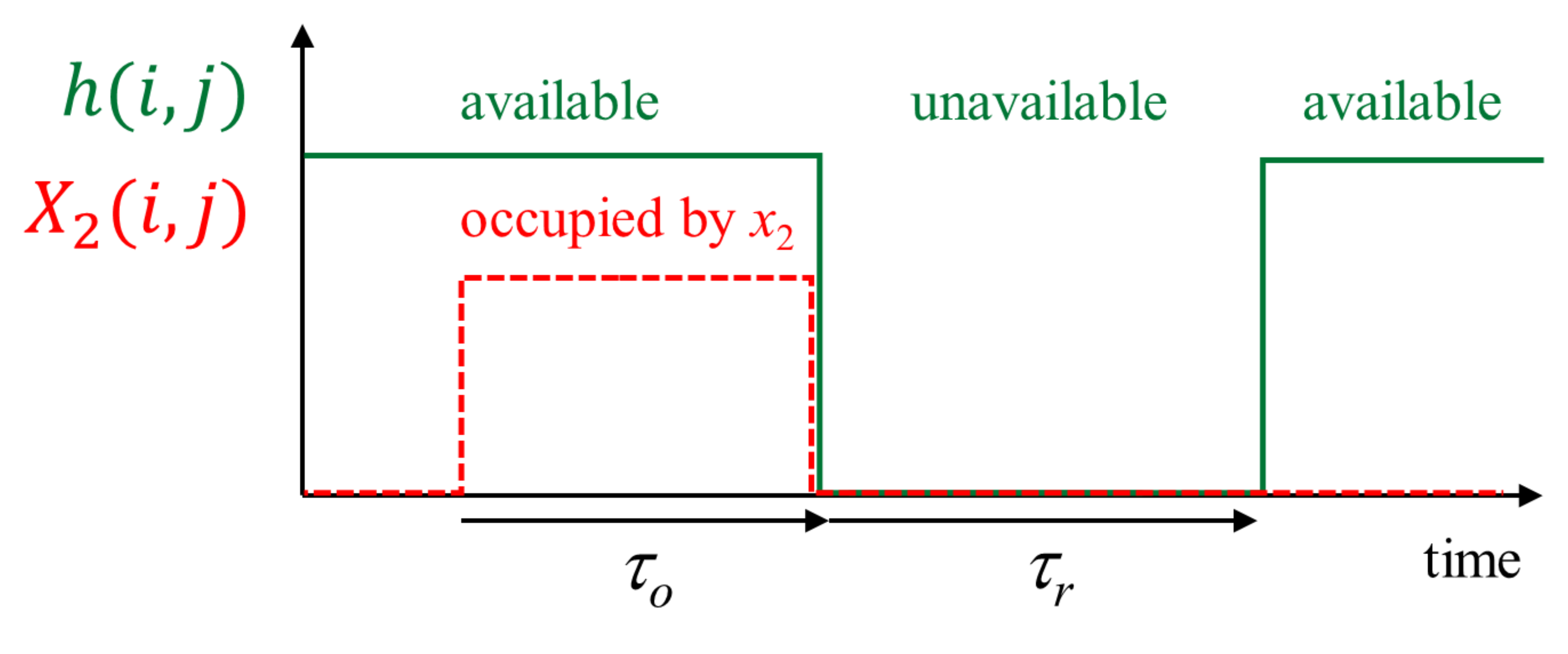}
 \caption{Dynamics of patch $(i,j)$. There are two characteristics times: the 
occupation time by the species $2$ until the patch becomes unavailable 
($\tau_o$), and 
the recovery time $\tau_r$.}
 \label{Figclock}
\end{figure}

We have implemented the dynamics described above in a simplified way, with a 
single resource $h$. We use two time scales, characteristic of the depletion by 
overgrazing ($\tau_o$) and the recovery of the resource ($\tau_r$). An internal 
clock at each patch takes care of the state of $h$. It starts counting at the 
colonization by the inferior competitor, and after a lapse $\tau_o$ of 
continuous occupation the resource is set to zero, local populations get 
extinct and the patch is flagged as unavailable for colonization 
(``desertification''). After a time $\tau_r$ the resource is 
recovered, and the patch is again suitable for colonization. 
Figure~\ref{Figclock} shows a schematic representation of this dynamics.

\subsubsection*{Numerical implementation}

In order to understand the effects of the state of the habitat on the 
occupation by the species, a spatially extended metapopulation model was 
implemented. The trophic web is embedded in a space structured like a square 
grid of side $L$. The borders of the grid are set 
as unsuitable for occupation, meaning that border conditions are closed. The 
possible states of each patch are: suitable or unsuitable for occupation ($h=1$ 
or $0$ respectively). 
The patches are initially all set at the $h=1$ state and both species occupy 
them in the same proportion and randomly distributed. 
If a population inhabits a patch, the variable of occupation $X_k(i,j)$ takes 
the value $1$ (here $k\in\{1,2\}$ is an index for the species, and $0< i,j <L$ 
are indices of position of the patch in the grid). If the species does not 
occupy the patch, $X_k(i,j)=0$. Because of the dependence of the state of the 
patches on previous occupation, a history is necessary at the start of the 
simulation. We also generate this at random, since the 
dynamics eventually overcomes any transient state. Time runs in a discrete 
way and at each time step the ecological processes occur in a stochastic way.

We run numerical simulations during a number of iterations that guarantees that 
the system reaches a stationary state in which the fraction of patches occupied 
by species $k$, defined as $x_k = \frac{1}{L^2} \sum_{i,j} X_k(i,j)$, 
fluctuates around a well defined value. In all the simulations we use $L=100$.

\section{Results}
In this section we present the main results obtained with the numerical 
implementation of the stochastic and spatially extended version of the model.  
We will leave for the next section the description of the results of the mean 
field model.

\subsubsection*{Evolution of the metapopulations}

\begin{figure}[t]
 \centering
 \includegraphics[scale=0.4]{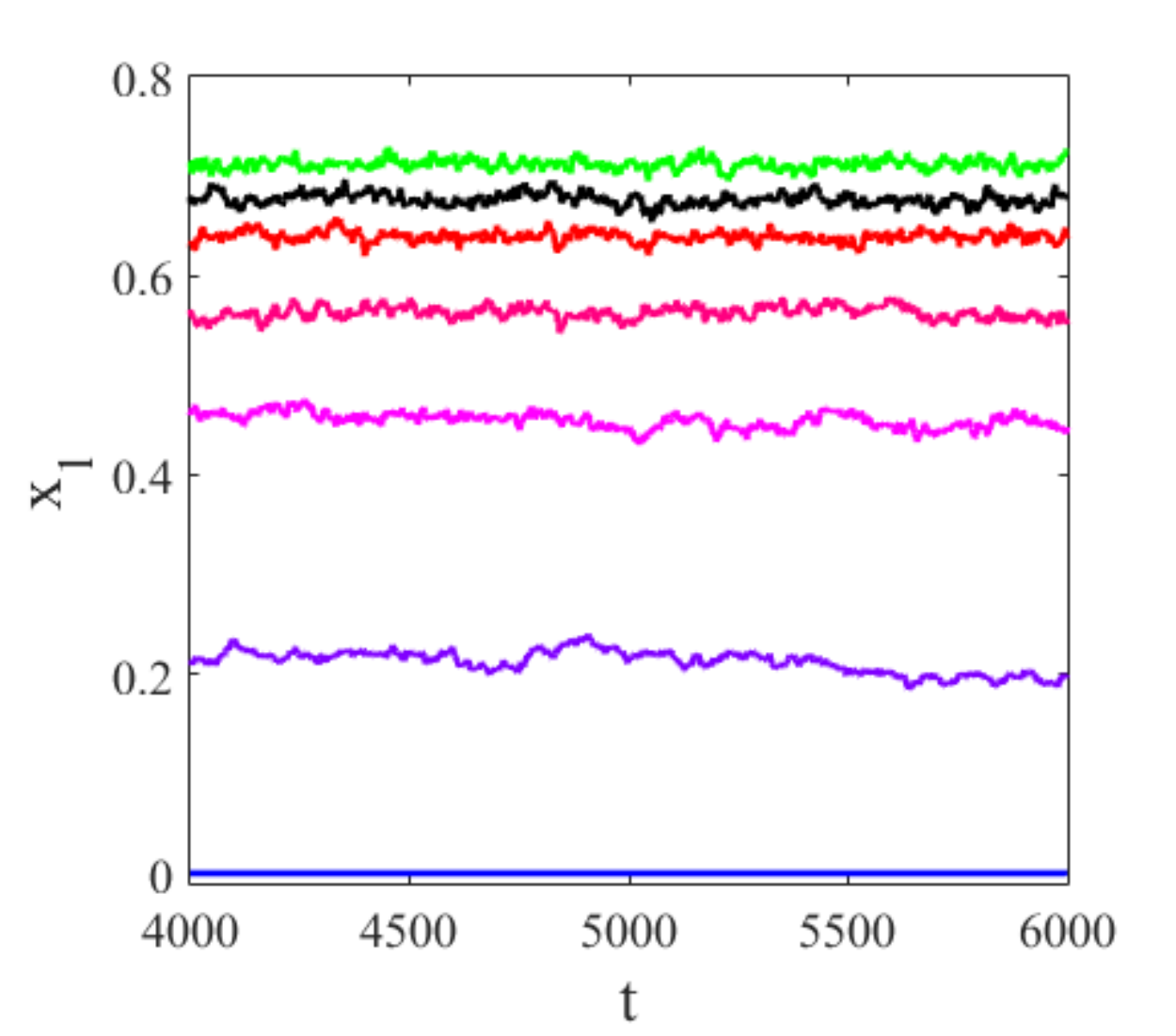}
 \includegraphics[scale=0.4]{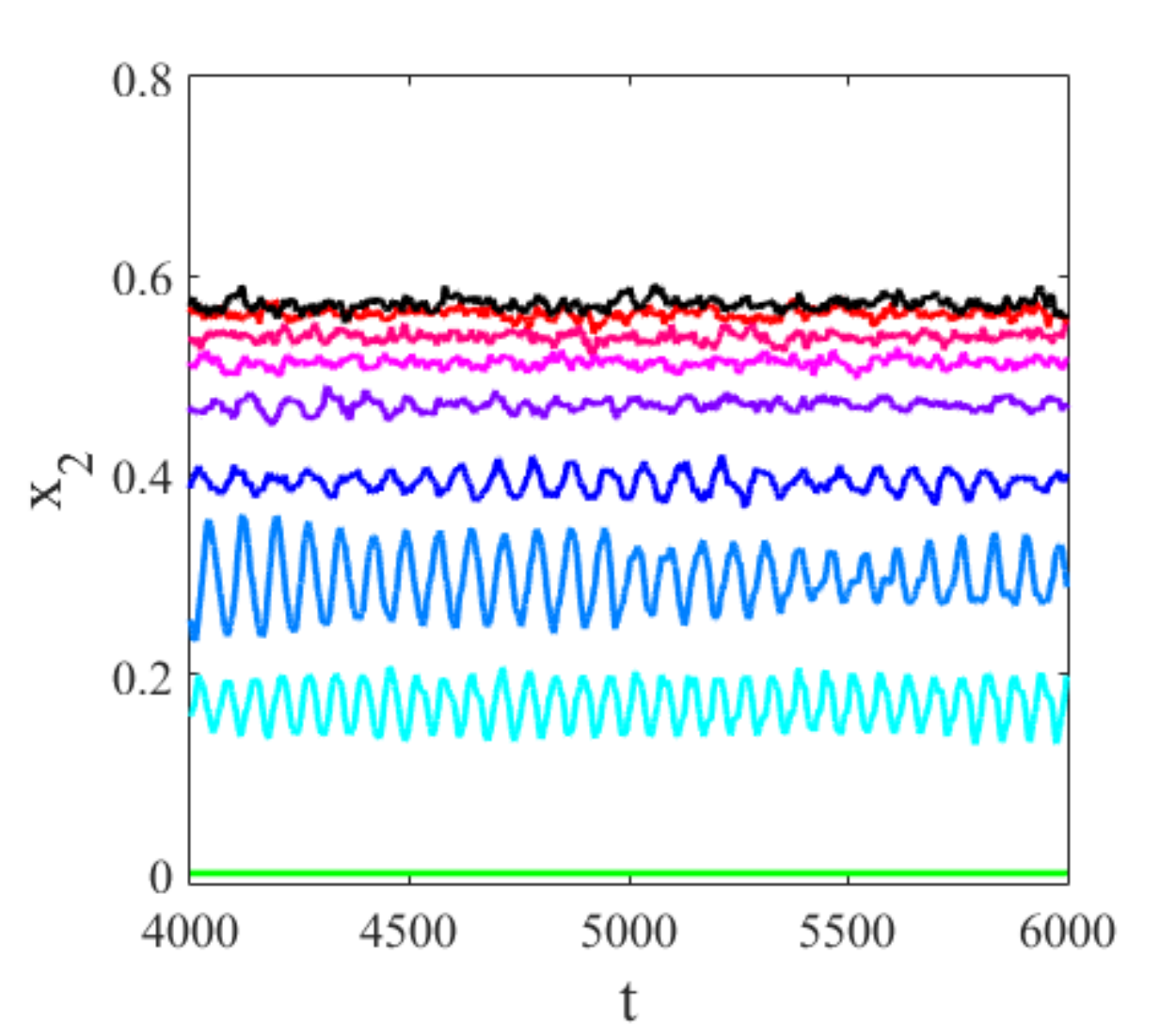}
 \includegraphics[scale=0.4]{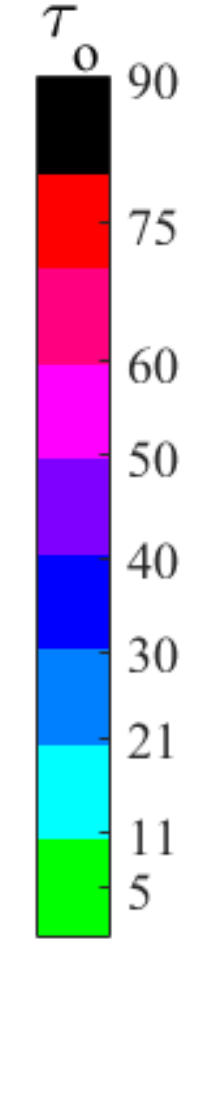}
 \caption{Temporal evolution of the fraction of patches occupied by the superior species 
($x_1$, left panel) and the inferior one ($x_2$, right panel) for several 
values of occupation time $\tau_o$. 
Colors indicate the value of $\tau_o$, going from cold (for low values of 
$\tau_o$), to warm ones (for high values of $\tau_o$). 
Parameters used: $c_{1}=0.05$, $c_{2}=0.7$, $e_{1}=0.05$, $e_{2}=0.01$, 
$\tau_r=50$.}
 \label{Fig1}
\end{figure}

Figure~\ref{Fig1} shows the behavior of typical populations of the two 
herbivores for several values of the occupation time $\tau_o$. The remaining 
parameters, $c_{1}$, $c_{2}$, $e_{1}$, $e_{2}$ and $\tau_r$, were held fixed at 
the values indicated in the caption of the figure.
In the graphs we plot the fraction of patches occupied by both species ($x_1$ 
for the superior herbivore and $x_2$ for the inferior one) at each time 
iteration. Each curve corresponds to a value 
of occupation time $\tau_o$. The color code goes from cold (small $\tau_o$) to 
warm colors (large $\tau_o$).

For small $\tau_o$  we observe a high level of occupation of the superior 
species ($x_1 \approx 0.7$ for $\tau_o=5$), which decreases abruptly  as 
$\tau_o$ increases; see the left panel of Fig.~\ref{Fig1}. In particular, for 
$11 \leq \tau_o \leq 40$ we obtain $x_1=0$, i.e. the superior species is 
extinct. When $\tau_o$ increases further, the fraction $x_1$ gradually grows, 
reaching values similar to those obtained for the smallest $\tau_o$ studied. 
In the right panel of Fig.~\ref{Fig1} we present the behavior of the inferior 
species. The fraction $x_2$ grows with increasing $\tau_o$ as the curves, from 
bottom to top, go from cold to warm colors. 
Moreover, we observe an oscillatory behavior for a range of values of the 
occupation time $\tau_o$, meaning that the patches are occupied and vacated by 
the $x_2$ population with a constant global period.

\begin{figure}[t]
 \centering
 \includegraphics[scale=0.29]{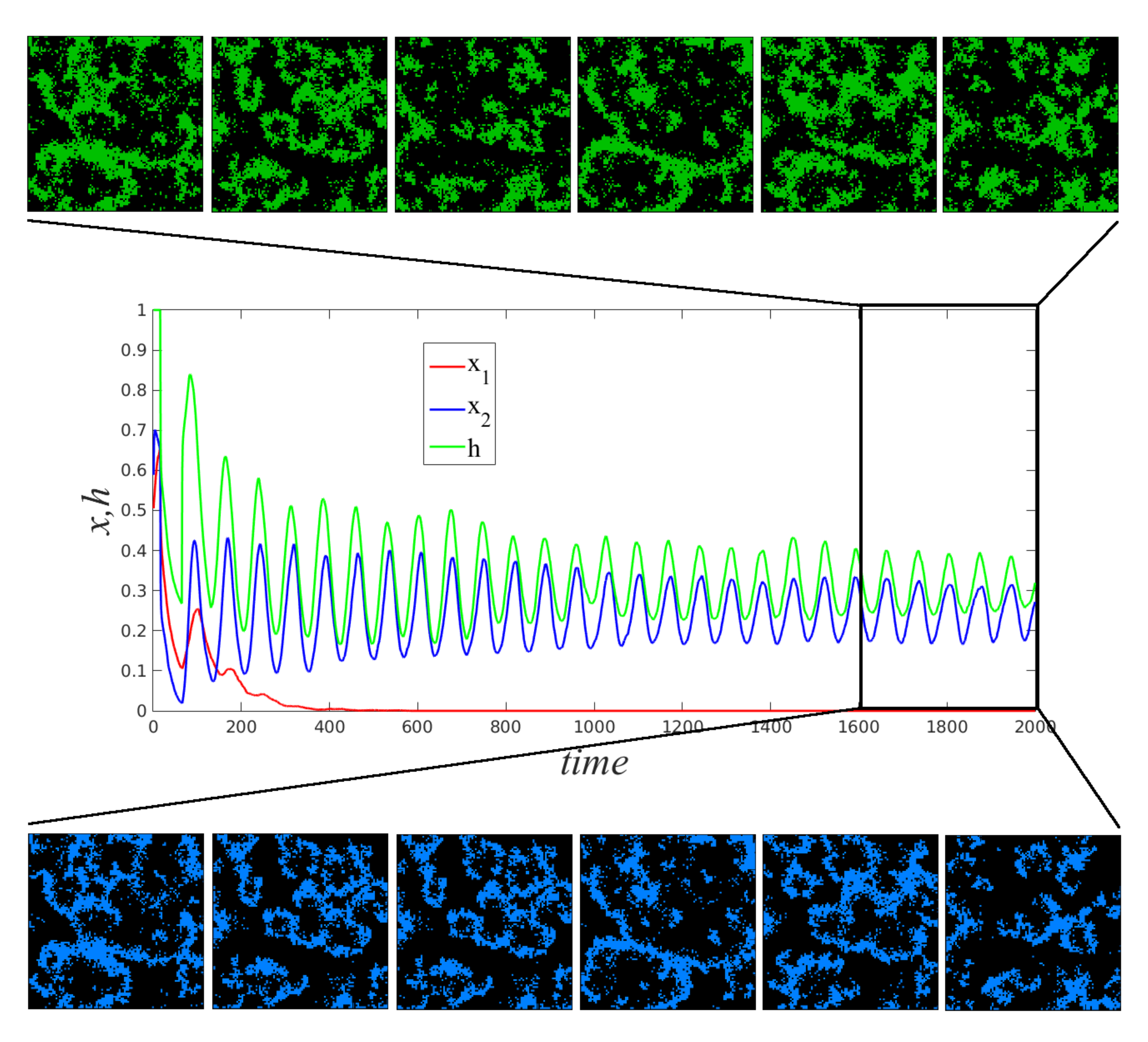}
 \caption{Top: Snapshots of the resource grid in the oscillatory regime. 
Available patches ($h=1$) are painted green. Center: Temporal evolution of 
the species and the resource for the same parameters. The inferior species 
oscillates with the same period as the resource, whereas the superior one was 
extinguished in all the patches ($x_1=0$). Bottom: Snapshots of the patches 
occupied by the inferior herbivore at the same time steps as the bottom panel. 
Patches occupied by $x_2$ are painted blue. Parameters used: $c_{1}=0.05$, $c_{2}=0.5$, 
$e_{1}=0.05$, $e_{2}=0.01$, $\tau_r=50$ and $\tau_o=17$.}
 \label{Fig2}
\end{figure}

Figure~\ref{Fig1} helps to draw the big picture of how the occupations evolve, 
and allows to illustrate three possible regimes in the dynamics of a given 
species, namely:
\begin{enumerate}
 \item extinction ($x_k=0$, i.e. emptying of all the patches),
 \item constant (non-oscillatory) survival ($x_k > 0$, fluctuating around a 
mean 
value but with no oscillations), and
\item oscillatory regime ($x_k > 0$, oscillating around a mean value with a 
well 
defined period).
\end{enumerate}
An exhaustive description of the oscillatory regime demands a visualization of 
the spatial grid in which the species are immersed.
We plot in Fig.~\ref{Fig2} some snapshots of the patches during the evolution 
of the system. It is clear that, even though the system only 
has local interactions (extinction, colonization and competitive 
exclusion involve at most a local neighborhood of a patch), a global dynamics 
of the system is attained by the propagation of waves of colonization and 
desertification. To complete the picture of this regime, the central panel of 
Fig.~\ref{Fig2} 
shows the fraction of patches occupied by the two herbivores 
together with the evolution of the resource. 
The existence of such periodical behavior suggests that we can 
characterize the different regimes by the frequency content of the occupations.

\subsubsection*{Characterization of the regimes}
\begin{figure}[t]
 \centering
 \includegraphics[scale=0.4]{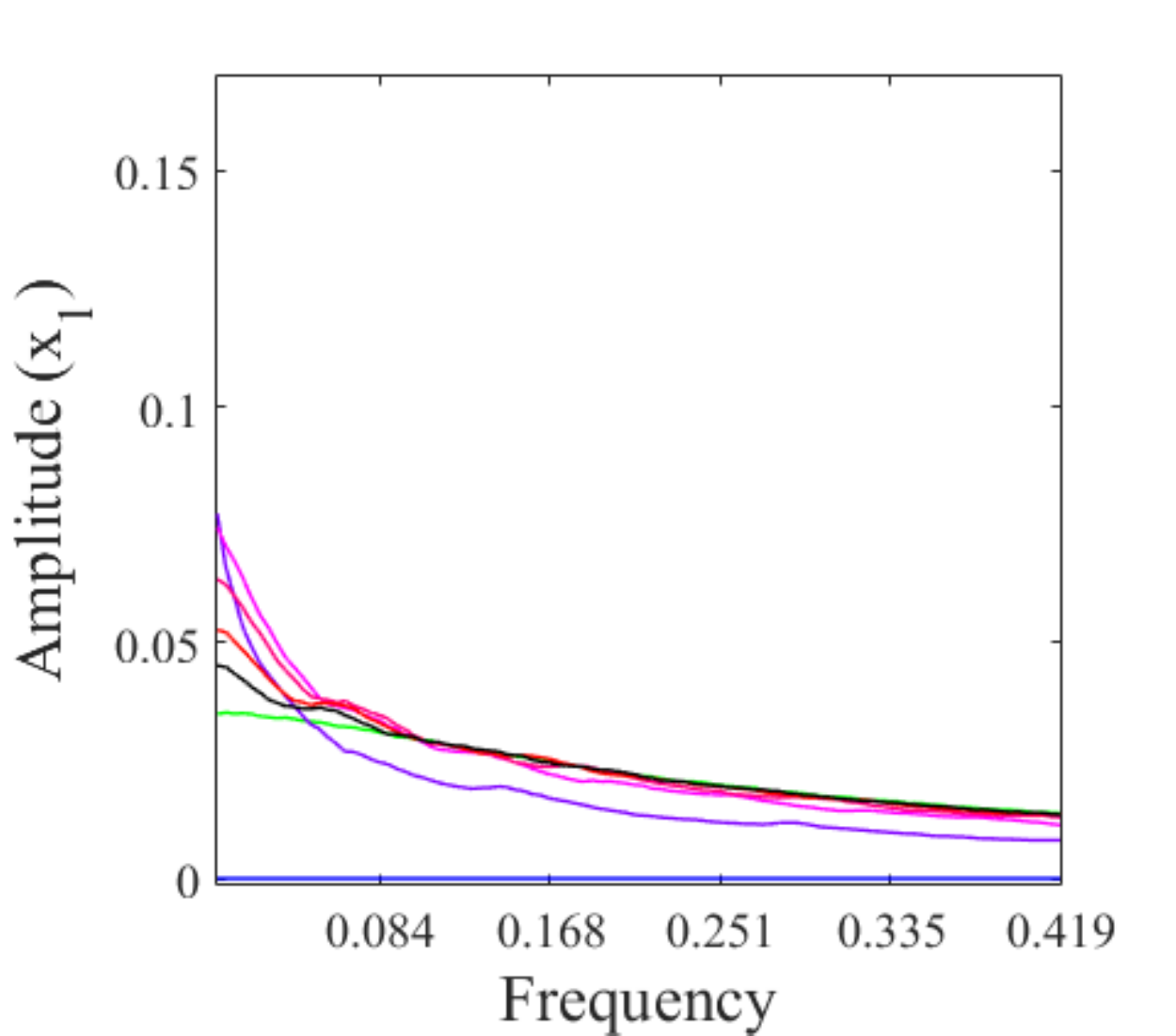}
 \includegraphics[scale=0.4]{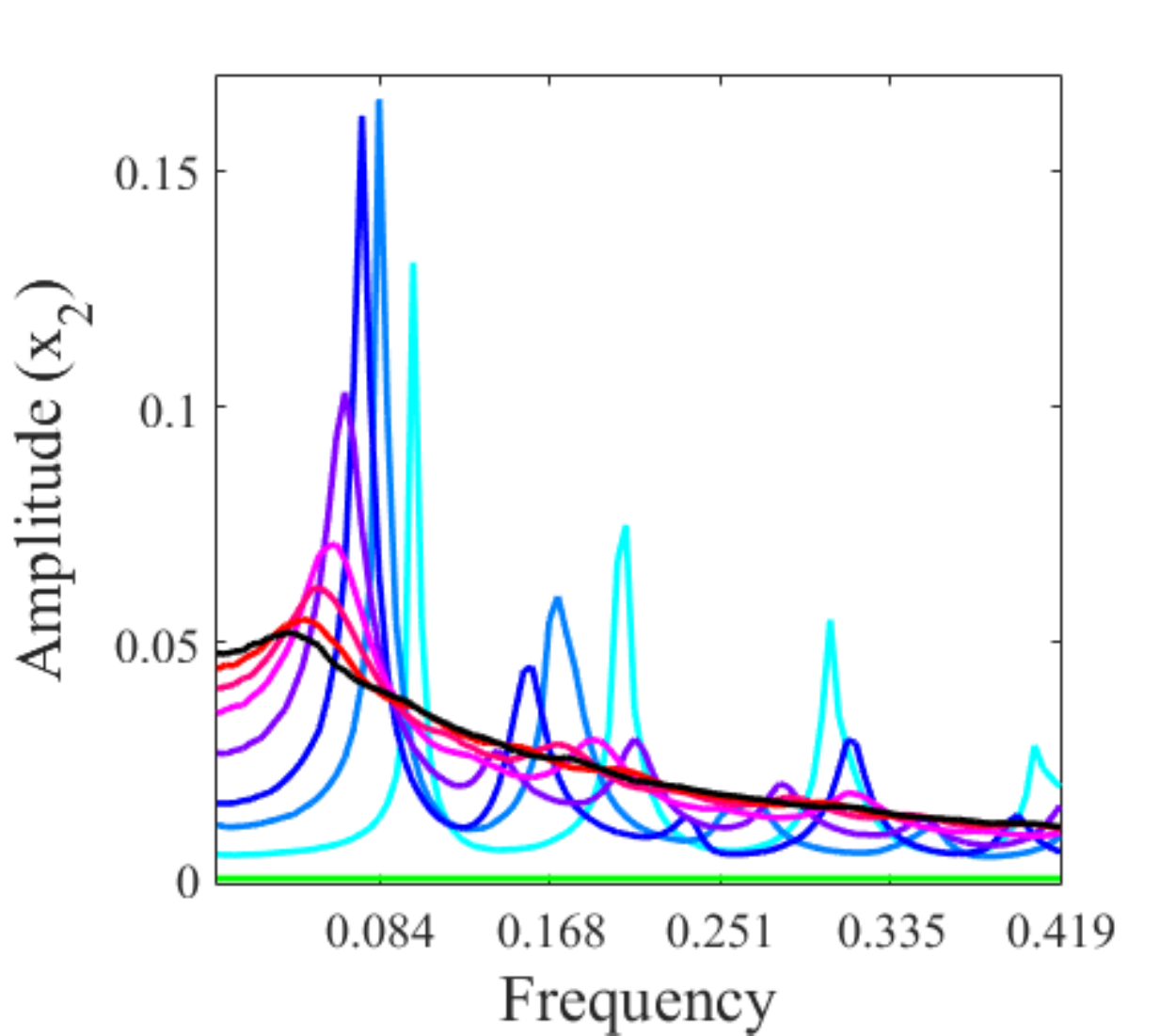}
 \includegraphics[scale=0.4]{barra_poblaciones3-eps-converted-to.pdf}
 \caption{ Left: Discrete Fourier transform of population $x_1$. Right: 
 Discrete Fourier transform of population $x_2$. Parameters used: 
$c_{1}=0.05$, $c_{2}=0.7$, $e_{1}=0.05$, $e_{2}=0.01$, $\tau_r=50$. Colors go  
from cold, corresponding to low values of $\tau_o$, to warm colors, 
corresponding to high values of $\tau_o$.}
 \label{Fig3}
\end{figure}

In order to determine the frequency spectrum of the time series of the 
occupations we calculate their Fourier transforms. 
The principal frequency (the most intense frequency of the spectrum), if 
present, indicates an oscillatory behavior of the system, meaning that the 
occupation is changing with a well defined period. In Fig.~\ref{Fig3} we plot 
the Fourier transforms of the curves of 
Fig.~\ref{Fig1}, calculated during the final $1500$ time iterations. In 
correspondence with Fig.~\ref{Fig1} the color scheme goes from cold colors 
(small $\tau_o$) to warm colors (large $\tau_o$). The spectra depict 
the presence of a periodicity in the occupations, even in the cases in 
which their evolution is rather noisy. In a wide region of the parameters space 
the spectra show the presence of a well defined frequency (and their harmonics) 
mounted on a basal signal originated in the stochastic noise intrinsic to the 
biological processes.

\begin{figure}[t]
 %\centering
 \includegraphics[scale=0.42]{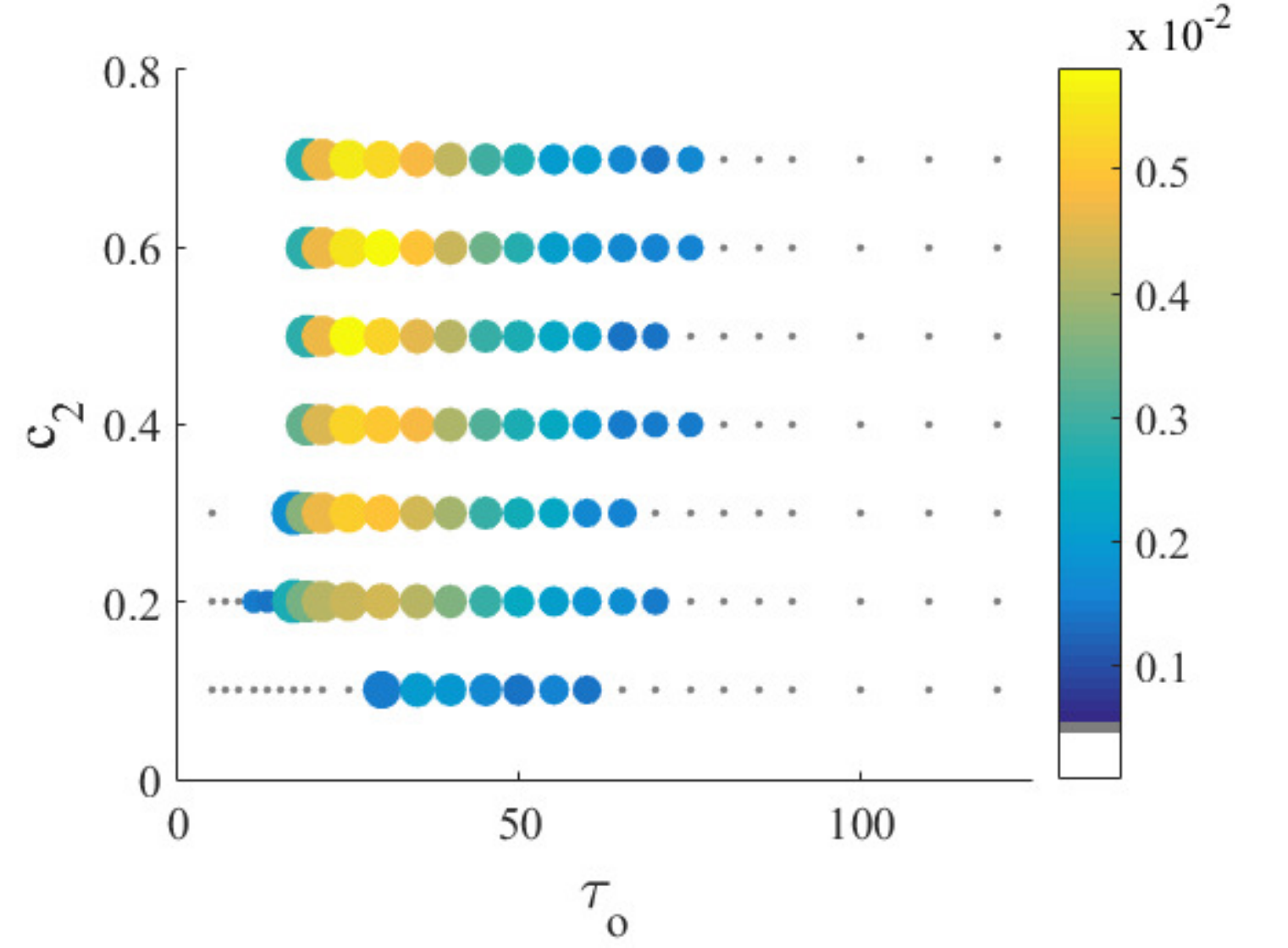}
 \includegraphics[scale=0.42]{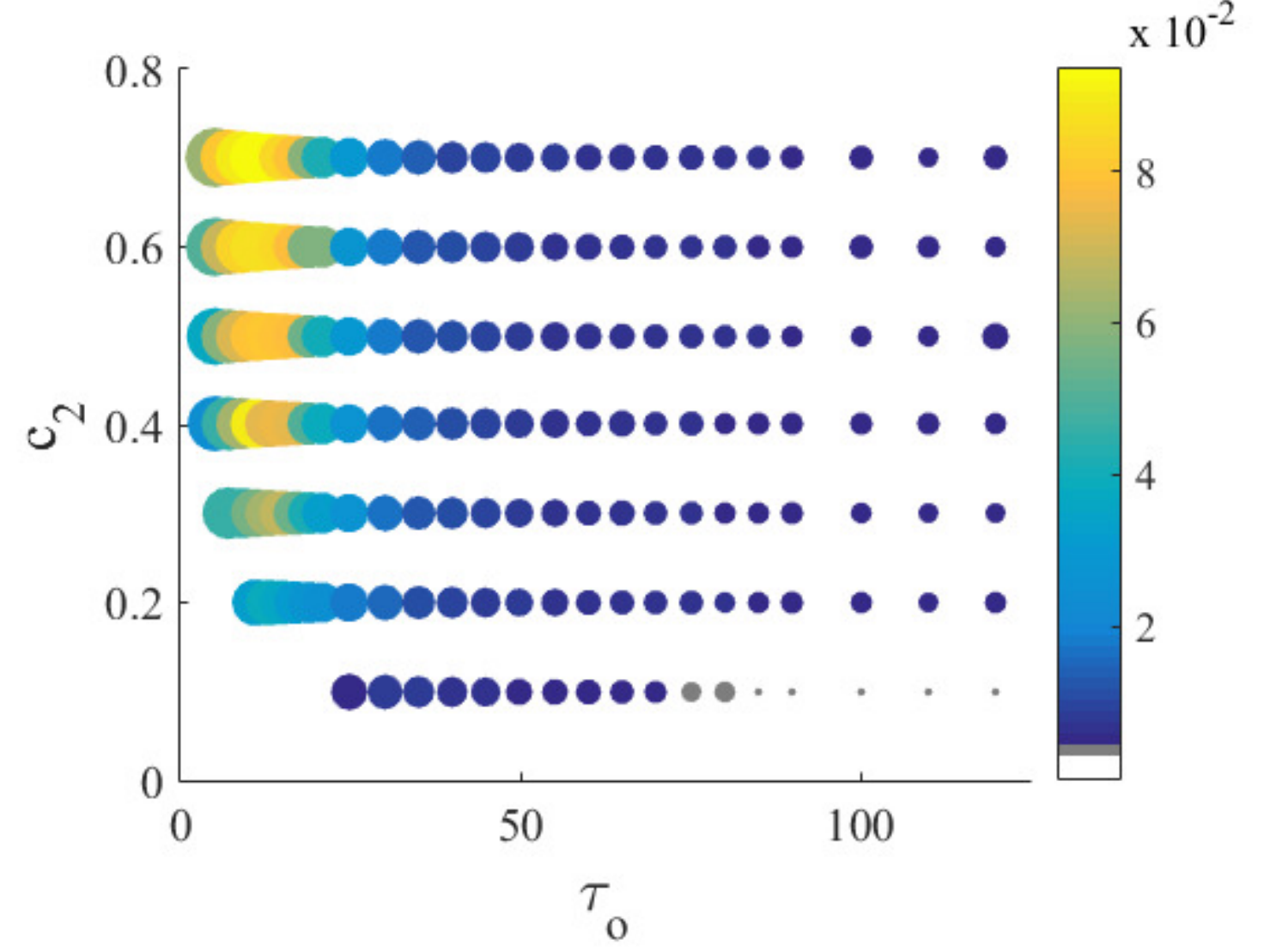}
 \includegraphics[scale=0.42]{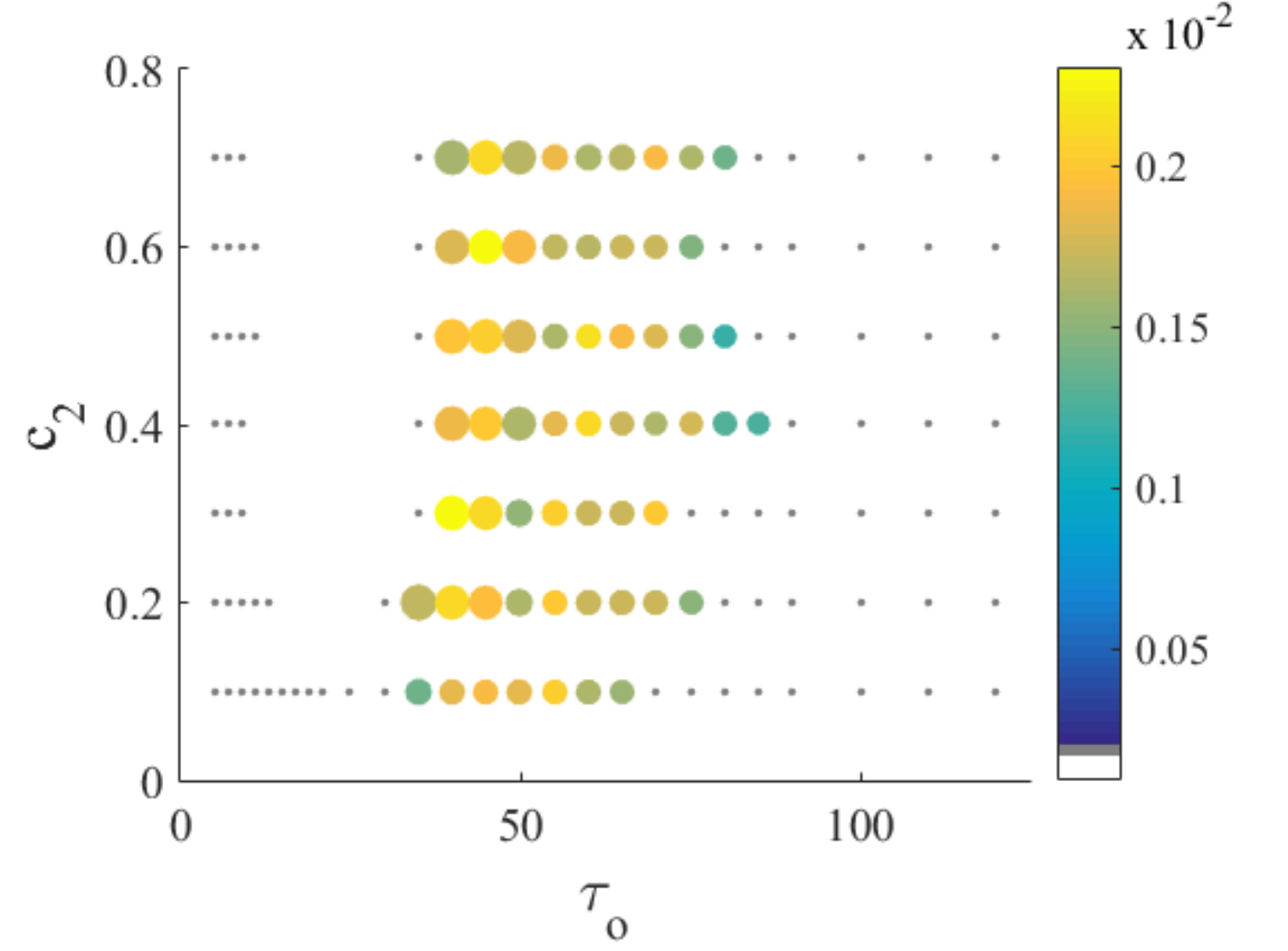}
 \includegraphics[scale=0.42]{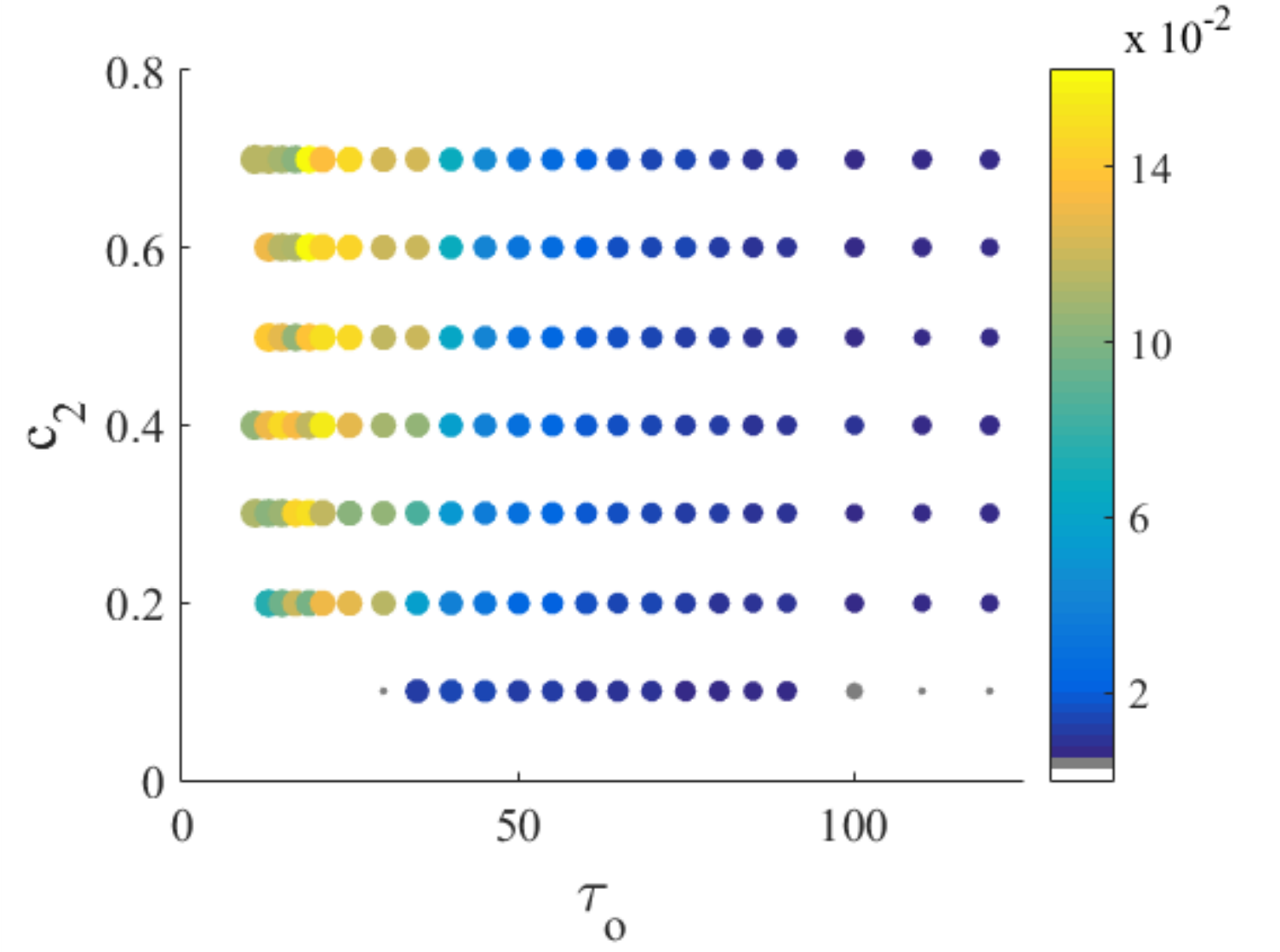}
 \caption{ Phase space for the parameters colonization rate $c_{2}$ and 
occupation time $\tau_o$ for a constant value of the patch recuperation time 
$\tau_r$. Left: $x_1$. Right: $x_2$. Top: $\tau_r=10$. 
Bottom: $\tau_r=50$. The size of the circles codifies the frequency and their color codifies the height of 
the maximum of the first significant peak. The absence of circles corresponds to extinction and small gray 
dots indicate non-oscillatory steady state. Please note that the color scales of $x_1$ and $x_2$, showing the height of the peaks, are not the same: for a given color, the peak of $x_1$ is smaller.}
 \label{Fig4}
\end{figure}

Although the detection of a principal peak in the Fourier transform 
indicates an oscillatory behavior, its presence 
is dubious in some cases because its intensity is very small and can not 
be separated from noise in a efficient way. 
For this reason we set up a phenomenological procedure that can be easily programmed to detect peaks automatically. From each spectrum we removed an exponentially fitted background, which leaves mainly the peaks if present. After this we run an automatic detection of peaks, and the spectrum was labelled as oscillating if it has a peak above a threshold. We set this threshold as 1\% of the amplitude of the largest peak for the corresponding parameters. This small value prevents overlooking the oscillations of $x_1$ even in the region where $x_2$ is synchronized with the resource (as in Fig.~\ref{Fig3}). It also allows a proper characterization of the decay of the amplitude of the oscillations as a function of $\tau_o$, as shown below.
With the information given by 
the first peak of the Fourier transform, we carried out a sweep in the phase 
space of the parameters of the model. The results obtained are presented if 
Fig.~\ref{Fig4} in the following way: for a constant value of recuperation time 
$\tau_r$, we measure the frequency and height of the maximum of the first 
significant peak. These features are codified in the size and color of the 
circles, respectively. Data are plotted as a function of the occupation time 
$\tau_o$ for each colonization rate $c_{2}$. 

Figure~\ref{Fig4} shows that, in all cases, the increase in the occupation time 
$\tau_o $ produces a decrease in the size of the circles (which encodes the 
frequency value of the main peak). That is, the frequency of occupation waves 
decreases with the occupation time $\tau_o $.
It is worth mentioning that the boundary between the zone of oscillations and 
the one of no oscillations is not abrupt but smooth, causing the border between 
these two regions to be blurred.
For the purposes of presenting the data, such zones are drawn taking into 
account the threshold described above.
It can also be observed that there are zones of extinction for either  
population, where the occupations of patches are null. The extinction regions 
move to large values of $\tau_o$ and they widen when the recuperation time of 
the patches $\tau_r$ increases.

Comparing the two top panels of Fig.~\ref{Fig4} it can be observed that in the 
zones where the population of the superior herbivore is extinct from all 
patches ($x_1=0$), $x_2$ has oscillations with large amplitude (lighter color). 
Moving to larger $\tau_o$, the amplitude of $x_1$ increases while the amplitude 
of $x_2$ decreases. 
Both amplitudes gradually decrease to low values in regions of large $\tau_o$.
Note also the difference in the scales of the amplitude of oscillations of $x_1$ and $x_2$. The presence of non-zero size circles and the scale of the intensity of the peaks in Fig.~\ref{Fig4} illustrates that there are oscillations in the $x_1$ population for intermediate values of $\tau_o$, with a very small intensity when compared with those of $x_2$; however, they can not be ignored.
The same can be said about the lower panels of Fig.~\ref{Fig4}, with the 
difference that the amplitudes change in a more homogeneous way in the zone 
post-extinction.

Figure~\ref{Fig4} confirm the existence of three dynamical regimes for each population. 
There are extinction zones where the populations disappear from the system, 
oscillation zones where the populations evolve forming traveling waves over 
the patches, and regions where the occupation reaches a non-oscillatory steady 
state. 
We will return to this phase diagram in the Discussion, where we will carry out 
a more detailed analysis of the different scenarios.

\section{Mean field approximation}

It is possible to formulate a deterministic mean field model which shares some 
similarities with the stochastic and spatially extended one defined above. In 
the spirit of 
the original Levins model of metapopulations and its generalizations 
\cite{levins1971,hanski1983} we propose the following delayed differential 
system, in 
terms of the density of patches $x_1(t)$ and $x_2(t)$, occupied by the superior 
and the inferior competitor respectively, and the available habitat $h(t)$:

\begin{align}
 \dot{x_1} &= c_1 (h - x_1 )x_1 - e_1 x_1  \label{eq:x1}\\
 \dot{x_2} &= c_2 (h - x_1 - x_2 )x_2 - e_2 x_2 \label{eq:x2}\\
 \dot{h}   &= -\gamma x_2(t - \tau_o )h(t - \tau_o ) + \gamma x_2(t - \tau_o - 
\tau_r )h(t - \tau_o - \tau_r ) \label{eq:h}
\end{align}

In this system, the parameters $c_1$, $e_1$ and $c_2$, $e_2$ represent the 
colonization and (local) extinction rates of the respective species. 
Equation~(\ref{eq:h}), 
representing the dynamics of the habitat, contains the delayed effect exerted 
by the species of the inferior competitor, responsible for the desertification. 
Observe that there is a negative contribution to the change of $h$, given by 
the occupied patches a time $\tau_o$ before $t$ (at a rate $\gamma$). There is 
also a 
positive one corresponding to the recovery of the destroyed habitat, after a 
time $\tau_o + \tau_r$ has elapsed.

It is important to stress that a model such as this ignores the short range 
correlations between occupied patches, which arise from the local and nearest 
neighbors dynamics of the extended system. This correlations play an important 
role in the most interesting behaviors of the spatial organization of the 
occupied space. Also, it ignores the role of fluctuations in the system. 
Nevertheless, some global features of the metapopulation dynamics can be 
captured by 
such an analytical model, and it is thus worth consideration.

\begin{figure}[t]
 \centering
 \includegraphics[width=\textwidth]{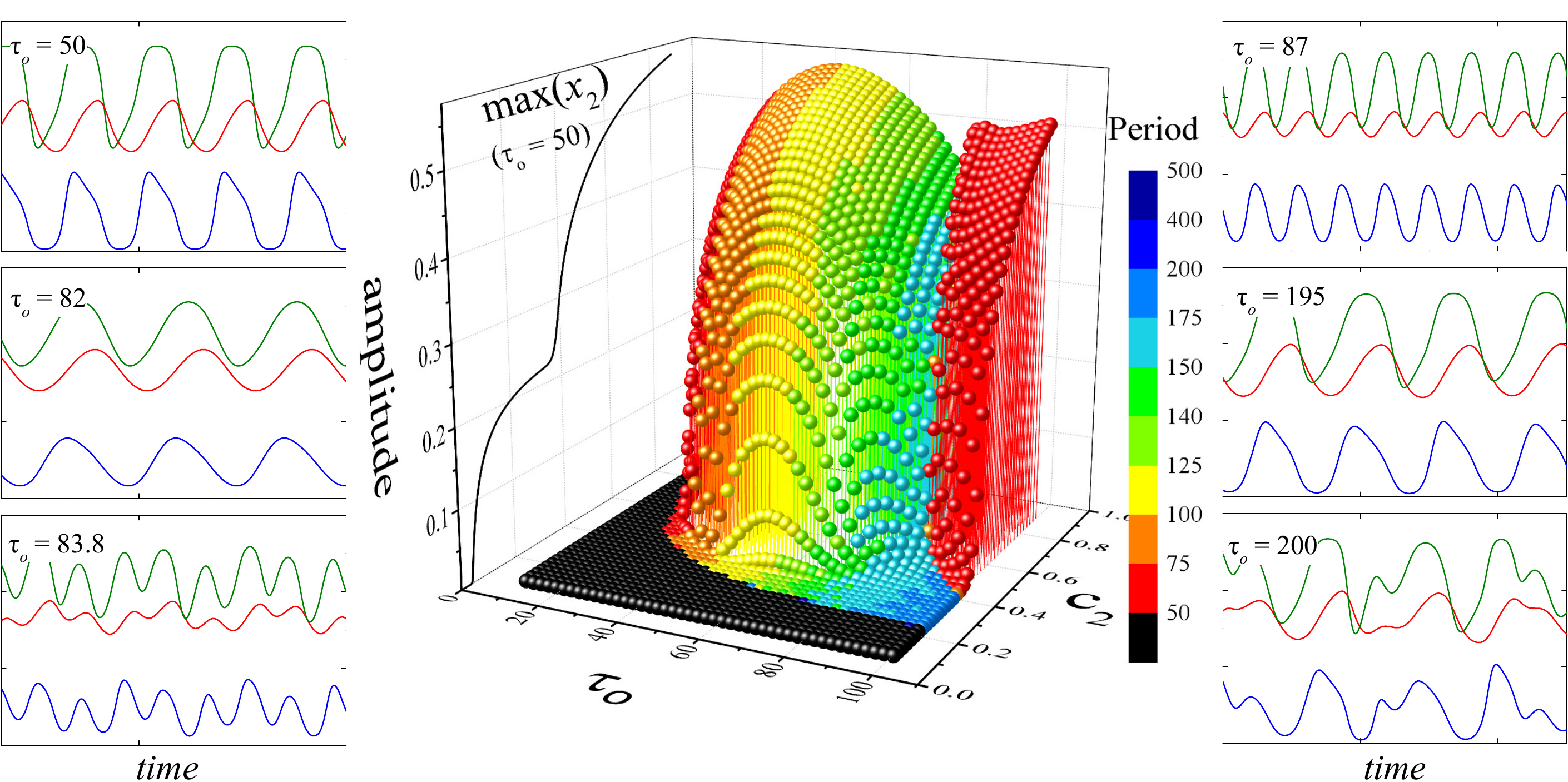}
 \caption{Phase diagram of the system (\ref{eq:x1}-\ref{eq:h}). Fixed 
parameters are: $\tau_r=10$, $c_1=0.04$, $c_2=0.7$, $e_1=e_2=0.01$, 
$\gamma=0.1$. The 
 lines show the evolution of $x_1$ (blue), $x_2$ (red) and $h$ (green).
 }
 \label{phases3d}
\end{figure}

We show in the central panel of Fig.~\ref{phases3d} a representation of the 
typical dynamical phases 
present in the system. The horizontal plane spans the parameters $c_2$ and 
$\tau_o$, 
while the rest of the parameters remain fixed. The vertical coordinate 
represents the amplitude of the $x_2(t)$ solution of the system after a steady 
state is 
achieved. Fixed points have amplitude zero and are represented by the flat 
floor of the graph. A line of Hopf bifurcations separates these fixed points 
from 
(three-dimensional) cycles, of which the amplitude of the $x_2$ variable is 
shown as height. These cycles are also characterized by their period, shown as 
a 
color scale. The leftmost wall of the plot shows a cut of this landscape at the 
intermediate value $\tau_o=50$. This curve represents the maximum of $x_2$, and 
serves the purpose of showing the distinction between two phases separated by a 
transcritical bifurcation in the fixed points region: the extinction of $x_2$ 
(a narrow band close to $c_2=0$) and a coexistence of the two species with a 
positive $x_2$. 

Besides these two bifurcations, there is a rich region of complex cycles 
attained by duplication of the period and chaos, separated from the simple 
limit cycles 
by a trough at $\tau\approx 80$ in Fig.~\ref{phases3d}. We have not explored 
this regime further, since in the present work our main interest is the 
behavior of 
the stochastic and spatially extended model, which more closely represents the 
situations found in real ecosystems. Some typical solutions are shown in the 
lateral panels. 
The period characterizing these phases (according to the colored 
scale) is the one with a strongest peak in Fourier space (which is just the 
period of the orbit in the case of simple cycles, before the period doubling), 
and the amplitude is the mean of the subcycles.

\begin{figure}[t]
 \centering
  \includegraphics[scale=0.4]{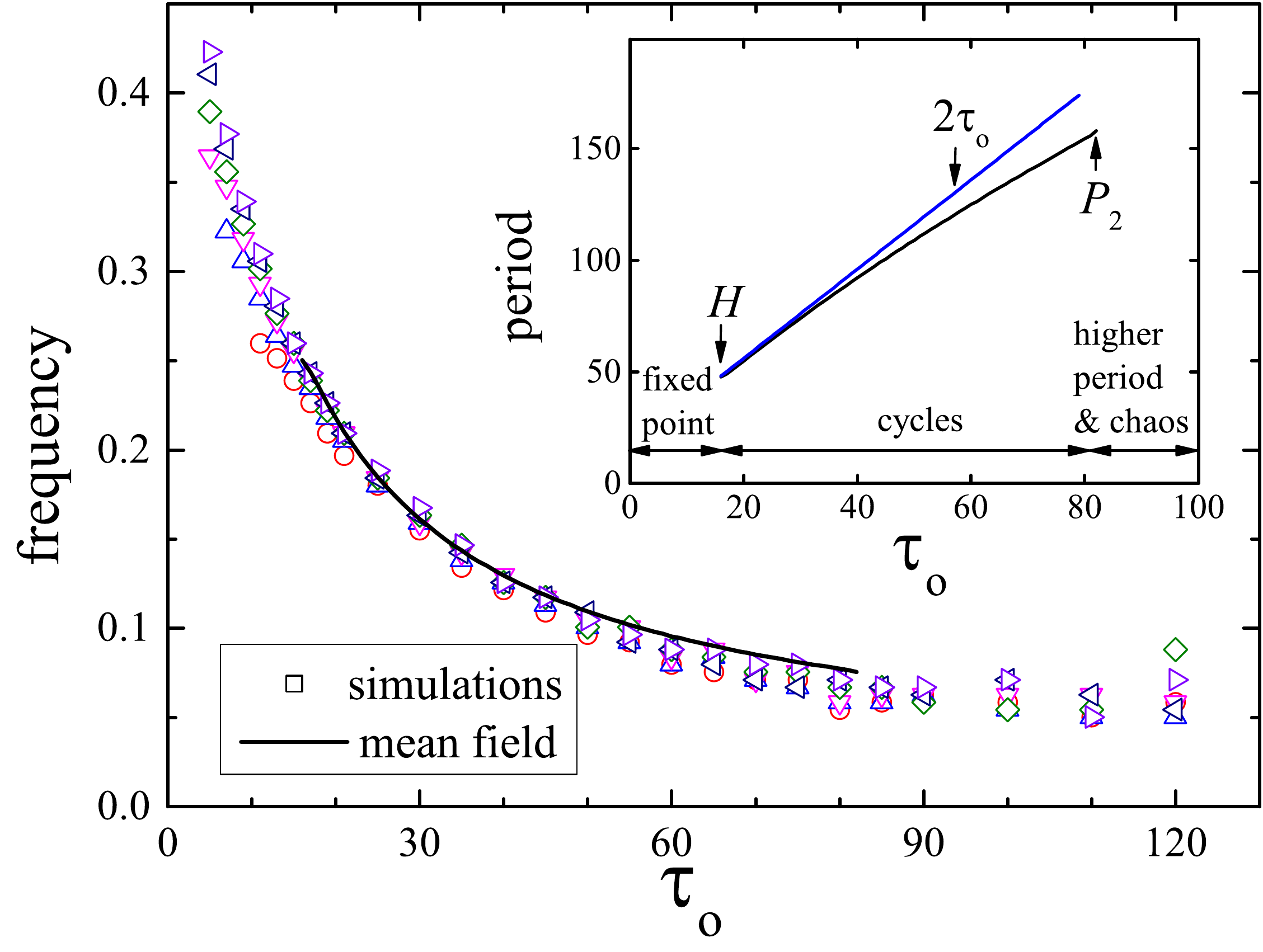}
 \caption{Diagram of the principal frequency of the occupation waves for $x_2$ 
in function of the occupation time $\tau_o$ for several values of colonization 
rate 
 $c_{2}$. The inset illustrates the analogue situation in the mean field 
approximation. In this case, the y axis indicates the period of the occupation 
waves.}
 \label{frecvstauo}
\end{figure}

As a final characterization of the oscillations displayed by the mean field 
model, we have plotted in the inset of Fig.~\ref{frecvstauo} the period of the 
cycles as a function of the 
characteristic time $\tau_o$, corresponding to the same parameters as in 
Fig.~\ref{phases3d} and $c_2$ fixed 
at $0.7$. The plot shows the range of simple limit cycles, from the point 
marked 
$H$ (the Hopf bifurcation) up to $P_2$ (the first period doubling bifurcation). 
After $P_2$ there exists the complex regime that we mentioned above, which is 
not displayed.

We can see that the 
period of the metapopulation solution grows with $\tau_o$---as it should---in 
the region between the Hopf bifurcation and the period doubling. It is 
remarkable 
that this dependence is not linear (a straight line of slope 2 is shown for 
comparison). In other words, increasing the characteristic time of 
desertification 
does not produce a proportional increase in the period of the population 
oscillation. A phenomenological allometric function of the form:
\begin{equation}
period = a+b\tau_o^c 
\end{equation}
gives an excellent fit to the measured period with $c=0.74\pm 0.04$, with an 
$R^2=0.99997$ confidence in the whole range of this phase. It is not easy to 
give a simple explanation for this departure from linearity. A sensible 
rationale behind it should consider that both phenomena, desertification and 
recovery, take place simultaneously in the system. While part of 
the system is 
becoming unavailable due to overgrazing, some other parts are already 
recovering from previous desertification events. The local population moves 
around the system (the metapopulation waves discussed above), allowing faster 
recovery of the occupation fractions, which is reflected in that the period 
bends down from a straight line as a function of $\tau_o$.

%\begin{figure}[t]
% \centering
% \includegraphics[scale=0.25]{periodo-vs-to-eps-converted-to.pdf}
% \caption{Period of the cycles between the Hopf bifurcation (indicated as $H$) 
%and the period doubling ($P_2$). A straight line with slope 2 is shown as a 
%reference. Fixed parameters are: $\tau_r=10$, $c_1=0.04$, $c_2=0.7$, 
%$e_1=e_2=0.01$, $o=0.1$.}
% \label{period}
%\end{figure}

The main plot of Fig.~\ref{frecvstauo} shows the corresponding frequencies of 
the stochastic simulation in the region of oscillations (reported in the 
previous section), also as function of $\tau_o$. Each set of points 
corresponds to a different value of $c_2$. We can see that the behavior is very 
similar to the one displayed by the mean field model (the black line). We have 
not observed the period doubling or the chaotic regimes in the simulations, 
most 
probably obscured by the fluctuations. Observe, nevertheless, an increase of 
the frequency (in some of the curves) for the largest values of $\tau_o$. Since 
these are the frequencies of the largest Fourier mode, this increase may be 
indicative of the regime of higher periods.

\section{Discussion}

In this work we have analyzed, in the framework of metapopulations, an 
ecological system inspired by a trophic web of two species of herbivores in 
hierarchical competition affecting an extended resource. 
For the populations of herbivores, three dynamical regimes were obtained in 
stochastic simulations as well as in a mean field approximation: extinction, 
oscillatory state, and non-oscillatory steady state.

The mean field model shows, for some values of the parameters, the existence of 
fixed points, corresponding to extinction or non-oscillatory behavior, and a 
the transition to a regime of cycles. These regimes are also observed in the 
stochastic model. The comparison between the two approaches can not be 
pushed too far, mainly because the mean field model does not take into account 
local interactions, spatial correlations and noise. Although it is expected 
that 
the long-range interactions 
inherent to the mean field contribute to synchronization, this is not the 
case for the stochastic (automata-like) model, with only local interactions. 
Therefore, it is remarkable that the synchronization of patches in the whole 
system is obtained in this case, by means of the drive imposed by 
the resource, which is in turn modulated by the behavior of the lower species 
$x_2$.
The noise inherent to the probabilistic character of the biological processes 
also 
difficult the comparison of the results of both models. 
The chaotic behavior could be present in the automata model, but it is probably 
hidden by the noise, since it has been observed in other ecological models  
that were analyzed from the point of view of coupled maps lattices.
In such systems the temporal evolution is obtained by means of maps that 
comprise terms of growing, intra- and interspecies competition, and coupling 
with local neighbors included as a diffusive operator \cite{soleetal1, 
soleetal2}. In that case the system shows the existence of chaotic 
attractors; spatial structures are formed even with the combination of 
parameters that gives chaotic dynamics \cite{soleetal1}.

\begin{figure}[t]
 %\centering
 \includegraphics[scale=0.25]{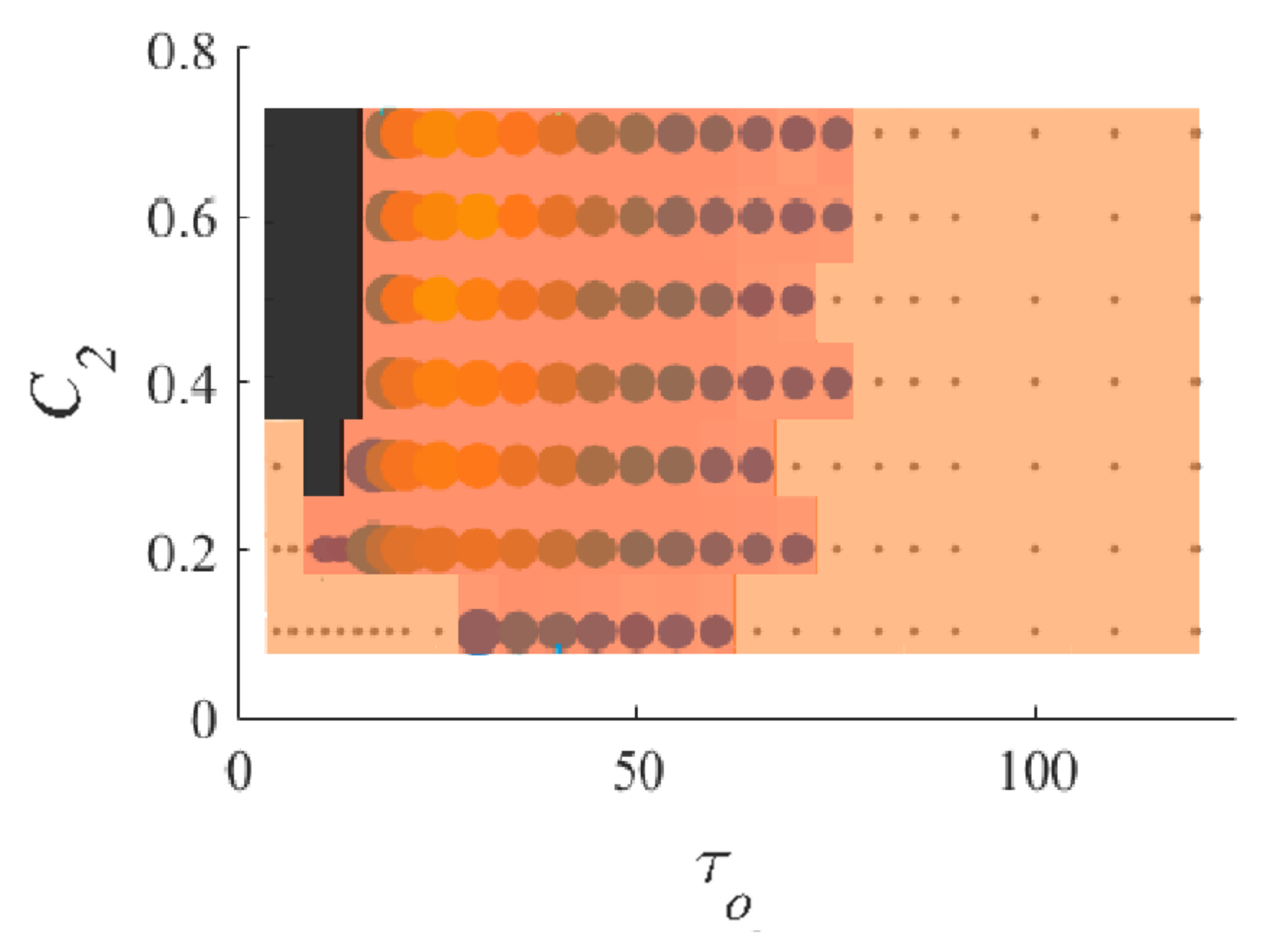}
 \includegraphics[scale=0.25]{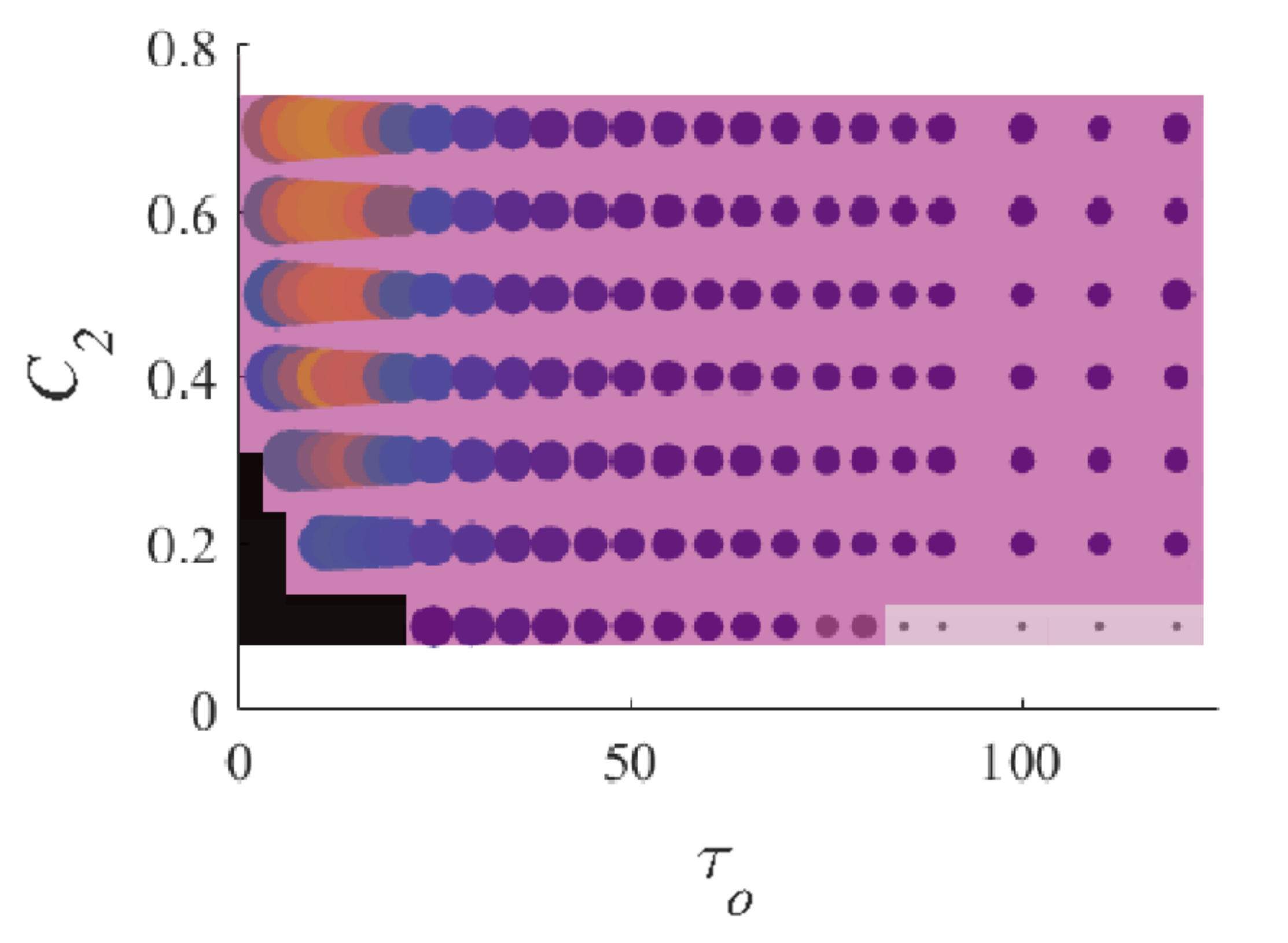}
 \includegraphics[scale=0.25]{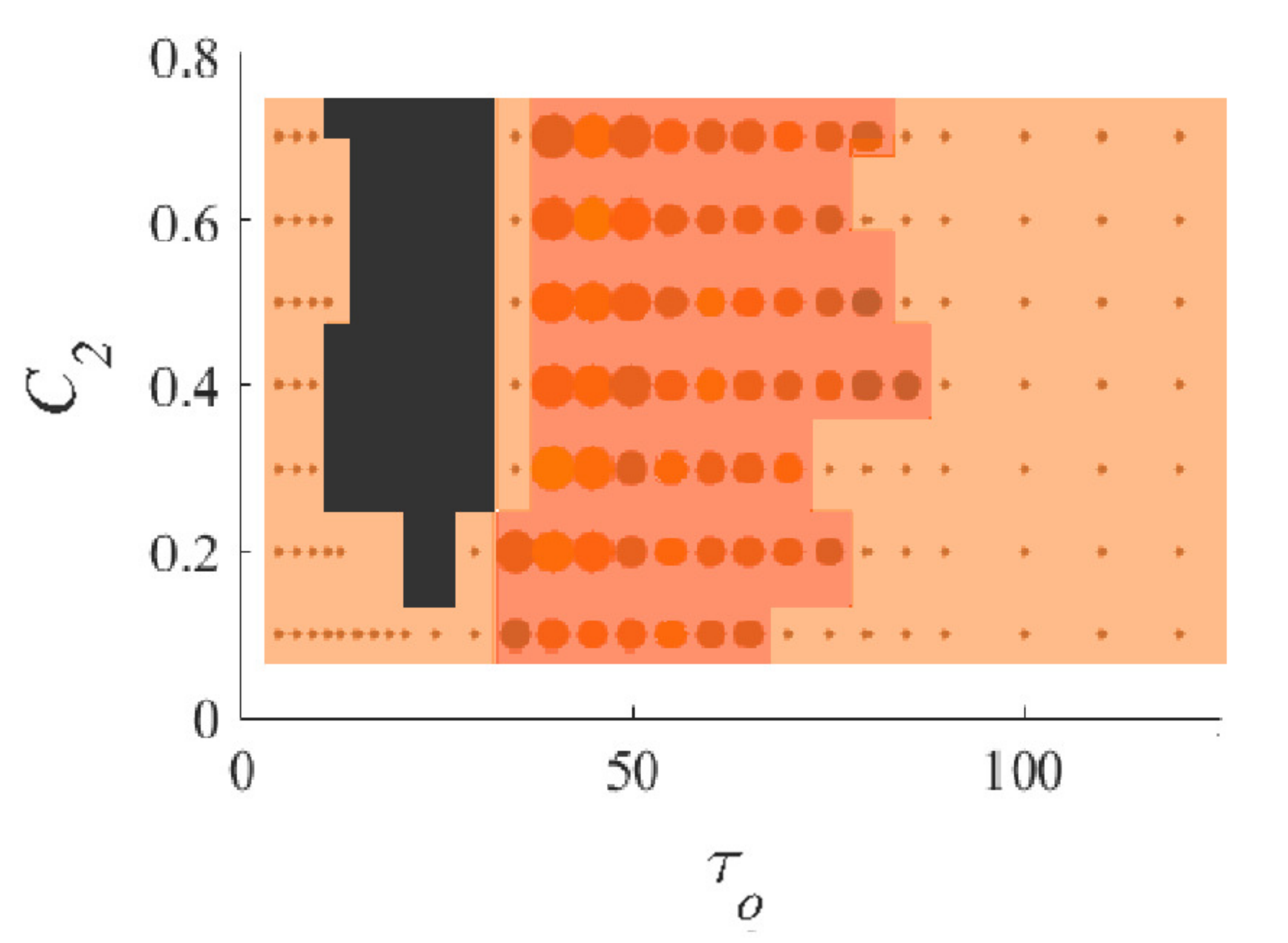}
 \hspace{0.5cm}
 \includegraphics[scale=0.25]{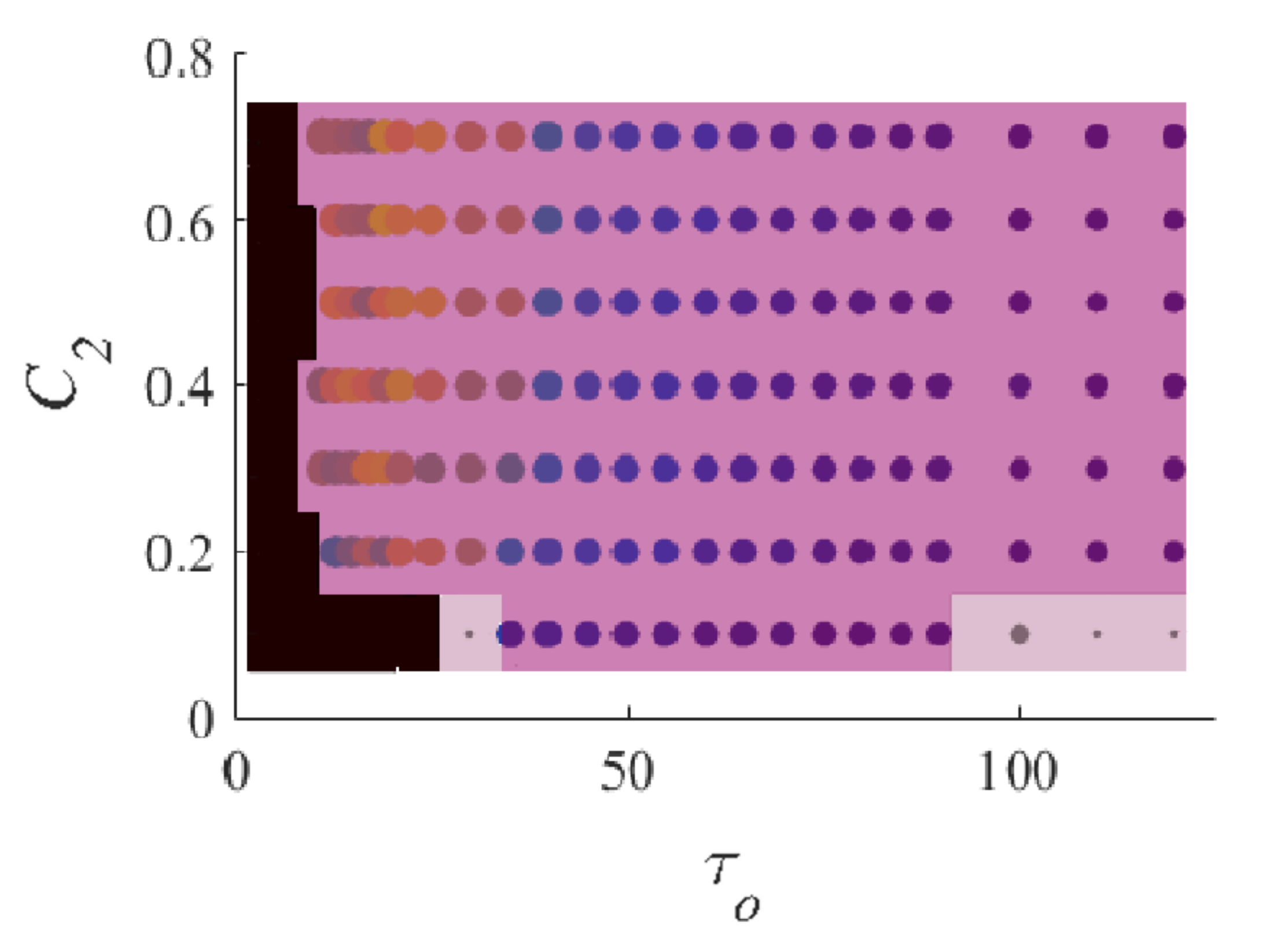}
 \caption{ Phase space of the parameters $c_{2}$ and $\tau_o$ (colonization rate 
 and occupation time) for a constant value of the patch 
recuperation time $\tau_r$. Left panels: $x_1$. Right panels: $x_2$. Top 
panels: $\tau_r=10$. Bottom panels: $\tau_r=50$. 
 Colors represent the regimes present in the dynamics: saturated colors for 
oscillatory regions, softer 
 colors for non-oscillatory steady state, and black for extinction.}
 \label{Fig5}
\end{figure}

In order to illustrate the observed dynamical behaviors, we plot in 
Fig.~\ref{Fig5} the phase space of Fig.~\ref{Fig4}, but now the 
different regions have been colored. Dark regions are the extinction zones. The 
remaining ones are colored with tones that differ in intensity, being more 
saturated in the zone of oscillations and less saturated in zones where the 
mean occupations come to a steady value (with no oscillations). 

By comparing the two top panels of Fig.~\ref{Fig5}, it can be seen that 
in the regions where $x_2=0$ (extinction of the inferior species), the 
population of the superior one $x_1$ comes to a non-oscillatory steady value. 
These results are obtained for low values of occupation time $\tau_o$, and low 
values of $c_2$, the capability of colonization of $x_2$, which make them unable 
to survive. Precisely, the absence of oscillations is due to the extinction of 
$x_2$, which in our model is the only one that affects the resource. That would 
be the state of the native species $x_1$ and the resource in the absence of the exotic species. 

The same can be said when the bottom panels of Fig.~\ref{Fig5} are compared. For both species 
the extinction zones enlarge when the recuperation time of the 
patch $\tau_r$ increases. 
Specifically, in the case of $x_2$, when the patches take more time to recover, 
even if $x_2$ is a better colonizer (a larger $c_{2}$), it is not enough to 
avoid extinction.
 
It can also be observed that the phase space has zones where both populations 
oscillate together, which occurs because the presence of $x_2$---directly 
coupled with the resource---induces oscillations in the patches that in turn 
produce oscillations in $x_1$. 
In fact, we can observe more clearly in Fig.~\ref{Fig4} that, for the 
parameters where $x_1$ is extinct, $x_2$ oscillates with a very 
large amplitude. In a real situation this could result into a very 
fluctuating income, which would not be
a desired behavior for a profitable management. 

There is also a region where the $x_2$ population does not exhibit oscillations 
for large values of the occupation time $\tau_o$. To understand what happens, 
observe the corresponding parameters there (lower right corner of the $x_2$ plots). 
The time $\tau_o$ can be seen 
alternatively as a measure of the environmental damage produced by the $x_2$ 
exotic population, because a longer $\tau_o$ means that $x_2$ is allowed a 
longer continuous presence at a patch (remember the patch comes to the $h=0$ 
state when the occupation time threshold $\tau_o$ is reached) and vice versa; 
then a longer $\tau_o$ results from a less harmful $x_2$. 
In other words, in this region $x_2$ cannot colonize so much (small $c_{2}$), 
but still survives because they are less harmful, plus they do not affect much the 
resource and then do not exhibit oscillations. 
Then, since the resource is not much affected by $x_2$, the native population 
$x_1$ can stay in a scenario near the natural conditions (without the exotic species), 
with long period oscillations of small amplitude. This is an unlikely scenario 
because it requires not so damaging herds.

In summary, if the occupation time $\tau_o$ is short, $x_2$ 
goes extinct if $c_{2}$ is small; but if $c_{2}$ increases, the population of 
$x_1$ becomes extinct. 
This situation gets worse if the recovery time, 
$\tau_r$, is longer. When $\tau_o$ is long 
the species coexist in a state where $x_1$ does not oscillate and 
$x_2$ show oscillations of low intensity and long period. 
For intermediate values of $\tau_o$, both species coexist 
exhibiting oscillations which increase their periods and decrease their 
intensities as either $\tau_o$ or $\tau_r$ increase.

\subsubsection*{Concluding remarks}

The mathematical model presented here reproduces the scientific evidence on the 
causes of desertification, whose main driving force 
is not natural conditions but due, almost purely, to the sum of local and 
short-term decisions \cite{andrade2013}. This transforming force is characterized by the predominance 
of ``I-here-now'' over the ecologically correct pathway in the long term 
\cite{monjeau2010} and has, as a consequence, a negative feedback between the 
sheep farming and the pastures that sustain it, causing the loss of 
productivity, biodiversity, quality of life; and after the collapse, the 
abandonment of a useless land.

In this context, our mathematical model can help decision makers see ``a movie''
that shows the catastrophic future that awaits them, unless they change their 
way of thinking and management of land use. 
%The lessons learned from such 
%a futuristic movie could even be applied to the whole of humanity and our 
%relation to natural capital, calling for a ``declaration of interdependence" 
%that we urgently need to assume in the era of the Anthropocene.

\section*{Acknowledgements}
We acknowledge financial support from several sources: Universidad Nacional de 
Cuyo (06/C506), ANPCyT (PICT-2014-1558) and CONICET (PIP 2015/0296).

%\bibliography{desertificacion}

\end{document}